\documentclass[12pt]{article}
\usepackage[dvips]{graphicx}
\def\sgn{\mathop{\mathgroup\symoperators{}sign}\nolimits}
\def\d{\,{\rm d}}
\begin{document}
\title{\vskip-40mm\rightline{\normalsize\bf KEK-CP-078}
\vfill
Spin and spin-spin correlations in\\
chargino pair production at future linear\\ $e^+e^-$ colliders
}
\author{\\
V.~Lafage$^1$, T.~Ishikawa$^1$, T.~Kaneko$^2$, T.~Kon$^3$,\\
Y.~Kurihara$^1$, H.~Tanaka$^4$\\ \\
$^1${\small\it High Energy Accelerator Research Organization (KEK)},\\
{\small\it Tsukuba, Ibaraki 305--0801, Japan\/}\\
$^2${\small\it Meiji-Gakuin University, Yokohama, Kanagawa 244--8539,
Japan\/}\\
$^3${\small\it Seikei University, Musashino, Tokyo 180--8633, Japan\/}\\
$^4${\small\it Rikkyo University, Toshima-ku, Tokyo 171--8501, Japan\/}\\
}
\date{}
\maketitle
\begin{abstract}
A possibility to measure the spin and spin-spin correlations of a
chargino pair is investigated in the process
$e^+e^- \to \tilde\chi^+_1 \tilde\chi^-_1 
\to$
$(\tilde\chi^0_1 q \bar{q}')$
$(\tilde\chi^0_1 q''\bar{q}''')$ 
at future linear-collider energies. 
The total and the differential cross sections are calculated by 
the {\tt GRACE} system which allows for the full spin correlation.
Experimental sensitivity of the measurements are examined by assuming
the limited detector resolution, the initial state radiation and 
the beam-beam effect (beamstrahlung). It is found that generally 
the spin-spin correlation can only be measured with a lower sensitivity 
than the chargino spin itself. The dependence of the correlation 
measurements on the relevant SUSY parameters can be seen for a light 
$\tilde\nu_e$ case, but the situation becomes worse for a heavier 
$\tilde{\nu}_e$. 
\end{abstract}
\newpage
\section{Introduction}
If Nature has chosen a supersymmetric scenario (SUSY) to build
the universe as many physicists expect, the chargino pair-production
process should be one of the first SUSY signals observed in the near
future at LEP2 or linear $e^+e^-$ colliders.
Charginos are mixed states of the spin-1/2 partners of the $W$ boson
and the charged Higgs boson,
and form two mass-eigenstates ($\tilde\chi^\pm_{1,2}$).
In many SUSY models, the lighter chargino, 
$\tilde\chi^\pm_1$, is thought to be the next-to-lightest SUSY 
particle while
the neutralino $\tilde\chi_1^0$
(a mixed state of SUSY partners of the photon, $Z$ boson and neutral
Higgs bosons), is the lightest SUSY particle. Then the main decay-mode
of the chargino is $\tilde\chi^\pm_1 \to \tilde\chi^0_1
f \bar{f}'$, where $f$ denotes a quark or a lepton.

Since the neutralino is invisible in the detector, the experimental
signature of the chargino pair-production is four jets or two jets and
one isolated lepton with large missing transverse momentum.
If the mass difference between chargino and neutralino is greater than
the $W$ boson mass, a real $W$ boson can be created and the two-body
decay, $\tilde\chi^\pm \to \tilde\chi^0 W$, should dominate.
In this case the additional signature of the $W$ boson production can
also be used for the event selection.
The mass of $\tilde\chi^\pm_1$ can be easily measured from a sharp
rise of the total cross section at its threshold.
The mass difference, 
$M_{\tilde\chi^\pm_1}-M_{\tilde\chi^0_1}$, 
can be measured from the maximum energy of 
$q\bar{q}'$ system~\cite{keisuke}.

If one wants to find other SUSY parameters beyond the measurements of
SUSY particle masses, one encounters some
difficulties\footnote{In ref.~\cite{keisuke}, they also pointed out
that the electron-beam polarization is helpful to separate the
higgsino component from gaugino one because the right-handed beam does
not couple to the gaugino component.}.
Since two invisible particles escape from the detection, a complete
reconstruction of the event kinematics is not possible. The same
situation happens in the $\tau$ pair-production: $\tau$ decays into
$\nu_\tau$ (invisible) and $q\bar{q}'$ system through virtual $W$.
After the discovery of $\tau$ lepton, it has been discussed~\cite{tau}
how to extract its weak properties. As a result it has been pointed
out that there are three angles which can be reconstructed
unambiguously from the experimental observables. They are
$\cos\theta^*_\pm$, and $\cos\Delta\phi^*$, where $\theta^*_\pm$ are
the polar angle of the $q\bar{q}'$ pair with respect to the momentum
of a mother particle ($\tilde\chi^\pm_1$ or $\tau^\pm$) measured in
the rest frame of the mother particle, and $\Delta\phi^*$ is the
azimuthal angle between the decay planes of the two mother particles.
This implies in turn that the spin and spin-spin correlations can be
measured from the limited number of experimental observables.

In order not to lose the precious azimuthal information, one must keep 
track of the decaying chargino polarisation (full spin correlation),
as done in~\cite{gmp} for application to angular distributions of
outcoming leptons. 

Recently Choi {\it et~al.\/}~\cite{choi} argued that the spin
information from the chargino decays directly reflects the chargino
mixing angles. They concluded that the chargino mixing angles and the
fundamental SUSY parameters could be determined from the measurement
of the total cross section of the chargino pair-production, their
spin, and spin-spin correlations.

In this report, we will discuss quantitatively the possibility to
measure those SUSY parameters from the spin and spin-spin correlations
in the chargino pair-production based on the detailed simulation which
also includes some detector effects and the smearing of colliding
energies due to the initial state radiation as well as the beam-beam
effect at future linear $e^+e^-$ colliders.

\section{SUSY parameters}
\subsection{Chargino description}
Charginos do not conserve fermion number.
Therefore the fermion number of charginos is not determined by
interactions but is a matter of convention.
In {\tt GRACE}~\cite{GRACE} we adopt the convention that the positively
charged charginos are Dirac-particles.
The two charginos are made of four Weyl spinors,
$\lambda^+$, $\lambda^-$, $\tilde H^-_1$ and $\tilde H^+_2$.
The corresponding physical states with mass $m_{\tilde\chi^\pm_1}$ and
$m_{\tilde\chi^\pm_2}$ are given by
\begin{equation}
 \Psi(\tilde\chi^+_i)=
 \left(\matrix{\overline{\lambda^-_{iR}}\cr\lambda^+_{iL}}\right),
\qquad
 \Psi(\tilde\chi^-_i)\equiv{\Psi(\tilde\chi^+_i)}^c=
 \left(\matrix{\overline{\lambda^+_{iL}}\cr\lambda^-_{iR}}\right),
\qquad
i=1,2
\label{eq:component}
\end{equation}
with
\begin{eqnarray}
\displaystyle
 \left(\matrix{\lambda^-_{1R}\cr\lambda^-_{2R}}\right)=&&
 \left(\matrix{\cos\phi_R&\sin\phi_R\cr-\sin\phi_R &\cos\phi_R}\right)
 \left(\matrix{\lambda^-\cr\tilde H^-_1}\right),\cr
\cr
\displaystyle
 \left(\matrix{\lambda^+_{1L}\cr\lambda^+_{2L}}\right)=&
 \left(\matrix{1&0\cr 0&\epsilon_L}\right)&
 \left(\matrix{\cos\phi_L&\sin\phi_L\cr-\sin\phi_L &\cos\phi_L}\right)
 \left(\matrix{\lambda^+ \cr \tilde H^+_2}\right).
\label{eq:mixing}
\end{eqnarray}
The two orthogonal matrices in (\ref{eq:mixing}) diagonalize the mass
matrix,
\begin{eqnarray}
 {\cal M}_C\equiv 
 \left(\matrix{M_2&\sqrt2M_W\cos\beta\cr\sqrt2M_W\sin\beta&\mu}\right)
\label{eq:mixmat}
\end{eqnarray} 
as
\begin{eqnarray}
  \left(\matrix{\cos\phi_L&\sin\phi_L\cr-\sin\phi_L&\cos\phi_L}\right)
 {\cal M}_C
  \left(\matrix{\cos\phi_R&-\sin\phi_R\cr\sin\phi_R&\cos\phi_R}\right)
 =\left(\matrix{m_{\tilde c_1} & 0\cr 0 & m_{\tilde c_2} }\right)
\label{eq:mixing_diag}
\end{eqnarray}
(the indexing of charginos is as follows:
$| m_{\tilde c_1}| < | m_{\tilde c_2}|$.) 

$\mu$ is the Higgs mixing parameter. In our convention, it is real and
can take any sign. It is one of the superpotential parameters.
As for $M_2$, it is a soft SUSY breaking parameter related to the mass
of the SU(2) gaugino.
The $\beta$ angle is related to the vacuum expectation values, $v_1$
and $v_2$, through
\begin{eqnarray}
 \tan\beta = {{v_2}\over{v_1}},\qquad
 \cos\beta = {{v_1}\over{\sqrt{v_1^2+v_2^2}}},\qquad
 \sin\beta = {{v_2}\over{\sqrt{v_1^2+v_2^2}}}.
\label{eq:vacuum_expectation}
\end{eqnarray}
The diagonal matrix with $\epsilon_L$ in (\ref{eq:mixing}) is to take
care of the possible negative eigenvalue for $m_{\tilde c_2}$.
(We can always choose the mixing parameters $\phi_R$ and $\phi_L$ such
that $m_{\tilde c_1}>0$). Practically, from (\ref{eq:mixmat}) we find
\begin{eqnarray}
\epsilon_L=\sgn(M_2\mu-M_W^2\sin2\beta)
          =\sgn\left(\det\left({\cal M}_C\right)\right).
\label{eq:SgnCharMass}
\end{eqnarray}
The physical masses of charginos are given by
\begin{eqnarray}
 m_{\tilde\chi_1^\pm}=            m_{\tilde c_1},\qquad
 m_{\tilde\chi_2^\pm}={\epsilon_L}m_{\tilde c_2},
\label{eq:char_phys_mass}
\end{eqnarray}
with $\tilde\chi^\pm_1$ lighter than $\tilde\chi^\pm_2$.

 The chargino spectrum is then related to the electroweak symmetry
breaking sector through $\mu$ and $\beta$, and to the SUSY
soft-breaking sector through $M_2$.

The couplings to $Z^0$ are determined by the sine and cosine of mixing
angles $\phi_R$ and $\phi_L$, as well as $\cos^2\theta_W$.

\subsection{SUSY spectrum evaluation}
Very few parameters are involved in the $2\to2$ chargino
pair-production process: besides those involved in the chargino
spectrum ($\mu$, $\beta$ and $M_2$), we only need the $t$-channel
related sneutrino $\tilde\nu_e$ mass ($m_{\tilde\nu_e}$).
However for the simulation of the experimentally more realistic
$2\to6$ process, a large part of the SUSY parameters must be known:
the neutralino spectrum and mixing matrix, the sfermions spectrum and
couplings.

 In order not to span the 80-dimensional SUSY parameters space, one
needs simplifying assumptions. The minimal supergravity (mSUGRA) is a
popular way to restrict this space.
mSUGRA keeps only 4 parameters: the common mass for scalar $m_0$, the
common mass for gaugino $M_{1\over2}$, the common (rescaled) trilinear
parameter $A_0$ and $\tan\beta$, as well as the possible choice of the
sign of $\mu$ ($\sgn(\mu)$), its absolute value being fixed by the
constraint of radiative electroweak symmetry breaking.
These parameters being determined at the grand unification scale
($\Lambda_{GUT}$), except $\tan\beta$, one needs to get their
running values at the electroweak scale ($M_Z$) through the
Renormalization Group Equation (RGE).
Further complications occurs due to the fact that this equation have
a two-boundary condition (some parameters are known at $M_Z$ and
others at $\Lambda_{GUT}$) and that the Yukawa's coupling are
not computed from the pole mass but from the running mass.
Some programs deal with all the detail of this computation. In this
study, the utility {\tt MUSE}~\cite{rge}, is used for this purpose.

In order to investigate the possibilities to measure the
gaugino-mixing angles, we tried to prepare three typical parameter
sets as proposed in~\cite{choi}:
gaugino-like ($M_2=81$~GeV, $\mu=-215$~GeV),
higgsino-like ($M_2=215$~GeV, $\mu=-81$~GeV)
and mixed ($M_2=92$~GeV, $\mu=-93$~GeV).
However, the constraints imposed by mSUGRA are quite restrictive.
The run down of $\tilde{m}^2_{H_2}$ (mass parameter of the
`up-sector' Higgs) is strongly pulled toward negative values by
$M_{1\over2}^2$. In order to get electroweak symmetry breaking, one
needs a large enough negative value for $\tilde{m}^2_{H_2}$.
So once $M_{1\over2}^2$ is fixed by the constraint on $M_2$, the only
remaining degree of freedom is $m_0$ ($A_0$ has not so much effect).

mSUGRA allows to get gaugino-like configurations, but not mixed or
higgsino-like.
Moreover once $m_0$ is fixed, $m_{\tilde\nu_e}$ is also determined.
Consequently, we chose an extended approximation with three common
scalar mass (one for the Higgs bosons, one for the squarks and one for
the sleptons), and we put $\mu$ `by hand' instead of computing it
with the radiative symmetry breaking constraint. The sneutrino mass
for the mixed scenario in the light sneutrino case had also to be put
by hand.
The resulting spectra are summarized in Table-1 and Table-2.

\section{Calculation method}
\subsection{Exact calculation}
All the cross sections given in this report, irrespective of various
levels of approximation used (see below), are numerically calculated
using the automatic calculation system {\tt GRACE}~\cite{GRACE}, based
on the helicity amplitude formalism. It includes the minimal SUSY
standard model~\cite{GRACE-SUSY}.
The cross section and any kinds of distributions can be obtained with
the aid of the multi-dimensional phase space integration package
{\tt BASES}~\cite{BASES}. 
Light fermion masses are also taken into account.

\begin{figure}[htb]
\centerline{
\includegraphics[width=\textwidth]{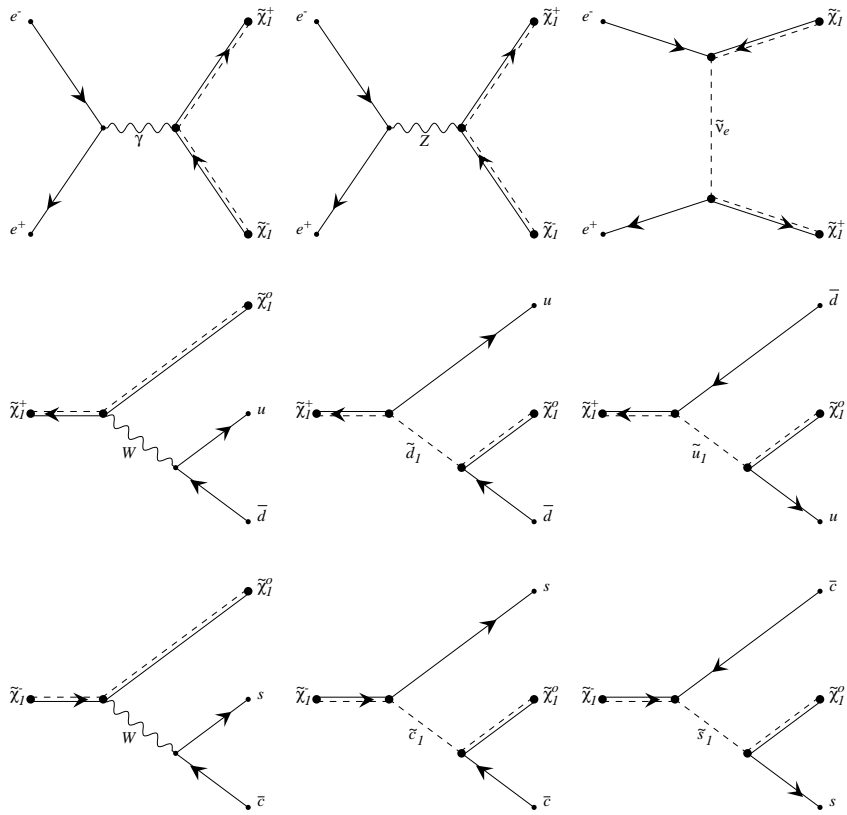}
}
\caption[FIG:diag]{\label{FIG:diag}
Feynman diagrams of a chargino pair-production (first row) and 
its decays (second row for 
$\tilde\chi^+_1$ and third row for $\tilde\chi^-_1$)
}
\end{figure}

The chargino pair-production in $e^+e^-$ collisions can be expressed
by the three Feynman diagrams shown in Fig.~\ref{FIG:diag} (first row).
It is followed by the subsequent decay of 
$\tilde\chi^\pm_1 \to \tilde\chi^0_1 q \bar{q}'$,
which has six diagrams in unitary gauge. 
In this report, the decay channels of 
$\tilde\chi^{+(-)}_1 \to \tilde\chi^0_1 u \bar{d} (s \bar{c})$
are chosen as benchmark processes.
Among those diagrams the contribution from Higgs-exchange and 
$\tilde{u}_2(\approx\tilde{u}_R)$ diagrams is found to be much smaller
than the statistical errors of the numerical integration (less than
0.5\%) and thus safely omitted.
Then three diagrams shown in Fig.~\ref{FIG:diag} (second row for
$\tilde\chi^+_1$ and third row for $\tilde\chi^-_1$)
are taken into account in our study. When we look at the final state of
$\tilde\chi^0_1 u \bar{d} \tilde\chi^0_1 s \bar{c}$
via the chargino pair-production, 
54 diagrams=$3\times3\times3\times2$ contribute to the process.
Besides those specified there remain a huge number of diagrams giving
the same final state but not through the chargino pair. 
Even with an automatic system like {\tt GRACE} the calculation with the
complete set of diagrams ($\sim$30,000) is hopeless within a reasonable
CPU time. To estimate the magnitude of the contribution from 
those background diagrams, we have calculated the cross section of 
the process $e^+e^- \to \tilde\chi^-_1 \tilde\chi^0_1 u \bar{d}$
with full diagrams. This process has 292 diagrams in unitary
gauge. Among them there are 9 diagrams which comes through the
chargino pair-production. The cross section above the threshold of
the pair-production has shown that the contribution from the background
diagrams was smaller than the statistical error of the numerical
integration. Hence it is confirmed that taking only the 54 diagrams
related to the chargino pair-production for the
$\tilde\chi^0_1 f \bar{f}' \tilde\chi^0_1 f \bar{f}'$
process is accurate enough up to the precision of 0.5\%.
It is worth mentioning that this calculation treating the amplitude
of the six-body final state can reproduce the full spin information
including the spin-spin correlation between charginos.

The decay width of the chargino are calculated by summing up five
possible decay channels with keeping fermion masses using {\tt GRACE}
system. Numerical results for the six parameter-sets are summarized in
Table.3. The total cross sections of six-particle final-state based on
the 54 diagrams are checked against those of the chargino
pair-production multiplied by the decay branching ratio at the center
of mass system (CMS) energy of 250~GeV. The results are consistent
one with each other as shown in Table.4 (first and second column).

To see the effect of the interference arising from the exchange of two
identical neutralinos, the cross section for 54 diagrams (considering
the statistical factor 1/2) is compared with that for 27 diagrams
omitting neutralino exchange. Both results completely agree within the
statistical error of the integration as shown in Table.4 (second and
third column).
This fact together with the other fact that the decay width of the
chargino is very narrow allows us to employ the narrow width
(on-shell) approximation to evaluate amplitudes of this process.

\subsection{Narrow width approximation with full spin correlation}
The simplest approximation of the full amplitude would be to take the
amplitude of the 2-body production process followed by the
{\em isotropic\/} decays. However, it is obvious that this
approximation is senseless when one talks about the spin measurement
as it ignores any spin information of the chargino and the vector
boson.
It is not a sufficient approximation even for the estimation of the
experimental acceptance of the detectors.

The best approximation to calculate the cascade decays of the SUSY 
particles should be as follows: connect the production amplitude for
$e^+ e^- \to \tilde\chi^+_1\tilde\chi^-_1 $ and the decay amplitudes
for $\tilde\chi^\pm_1 \to \tilde\chi^0_1 q \bar{q}'$ tracing the
helicity of each particle.
The cross sections is then expressed as
\begin{eqnarray}
\sigma&=&\frac{1}{C} \sum_{h_i} \int\!\!\!\int\!\!\!\int
{\left|\sum_{h_+,h_-}{\cal A}_{e^+e^-\to
\tilde\chi^+_1\tilde\chi^-_1} \cdot
{\cal A}_{\tilde\chi^+_1\to \tilde\chi^0_1 u \bar{d}}\cdot
{\cal A}_{\tilde\chi^-_1\to \tilde\chi^0_1 s \bar{c}}\right|}^2
\d\Omega_{2\to2}
\nonumber \\
&~&\times
\frac{\d\Omega_{1\to3}}{2m_{\tilde\chi^\pm_1}\Gamma_{\tilde\chi^\pm_1}}
\frac{\d\Omega_{1\to3}}{2m_{\tilde\chi^\pm_1}\Gamma_{\tilde\chi^\pm_1}}
\nonumber
\end{eqnarray}
where ${\cal A}_{e^+e^-\to \tilde\chi^+_1\tilde\chi^-_1}$ and 
${\cal A}_{\tilde\chi^-_1\to \tilde\chi^0_1 u(s) \bar{d}(\bar{c})}$
are respectively the amplitudes of the chargino pair-production 
and decay, 
$h_i$ the helicity for the initial and the final particles, 
$h_\pm$ the helicities of $\tilde\chi^\pm_1$,
$\d\Omega_{i\to j}$ the phase space for the production and the
decay, $m_{\tilde\chi^\pm_1}$ the mass of chargino,
$\Gamma_{\tilde\chi^\pm_1}$ the decay width of chargino, and
C = (spin average factor)$\times$(flux factor).
In this method the spin summation is taken not only over the diagonal
part but also over the off-diagonal one.
This is equivalent to deal with the full spin density matrix for the
chargino decay.

When integrating over the whole phase-space (without cuts), the
off-diagonal part disappears (but one must keep it when studying
differential distributions), and 
the total cross sections can be expressed as:
\begin{eqnarray}
\sigma&=&\sigma_{e^+e^-\to\tilde\chi^+_1\tilde\chi^-_1}
\cdot \frac{\Gamma_{\tilde\chi^+_1\to \tilde\chi^0_1 u \bar{d}}}
{\Gamma_{\tilde\chi^\pm_1}}
\cdot \frac{\Gamma_{\tilde\chi^-_1\to \tilde\chi^0_1 s \bar{c}}}
{\Gamma_{\tilde\chi^\pm_1}}
\nonumber \\
&=&\sigma_{e^+e^-\to\tilde\chi^+_1\tilde\chi^-_1}
\cdot Br(\tilde\chi^+_1\to\tilde\chi^0_1 u \bar{d})
\cdot Br(\tilde\chi^-_1\to\tilde\chi^0_1 s \bar{c}),
\nonumber 
\end{eqnarray}

\begin{figure}[htb]
\centerline{
\includegraphics[width=\textwidth]{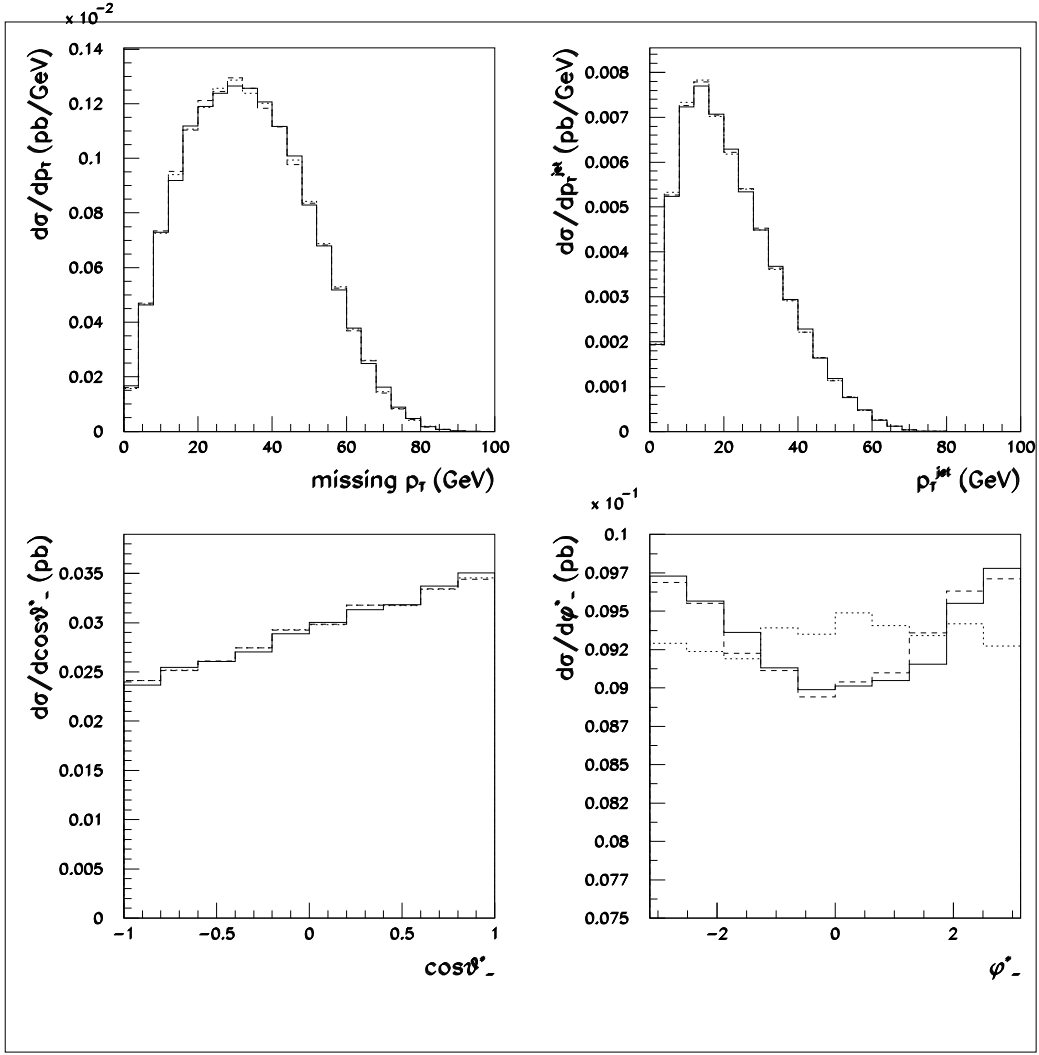}
}
\caption[FIG:comparison]{\label{FIG:comparison}
Distributions obtained from the exact calculations (solid lines),
narrow-width approximation with full spin-correlation (dashed lines),
and it with diagonal spin-correlation (dotted lines)
with the parameter set `gaugino region, light $\tilde{\nu}_e$' at
the CMS energy of 250~GeV.
The definitions of the variables are given in the text.
}
\end{figure}

\begin{figure}[htb]
\centerline{
\includegraphics[width=\textwidth]{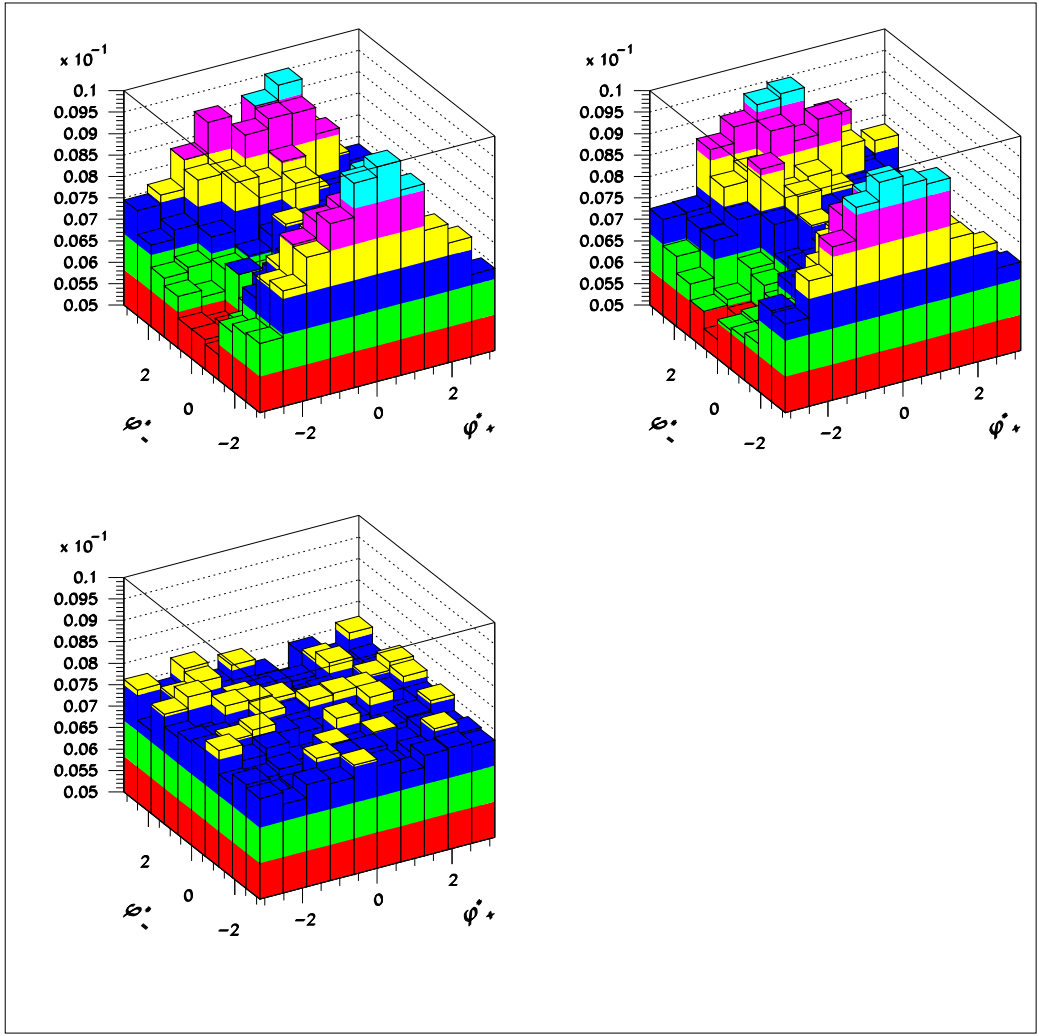}
}
\caption[FIG:phiphi]{\label{FIG:phiphi}
The correlation of azimuthal angles of the two decay planes of the
chargino decays obtained from (a) the 54 diagrams $exact$ calculation
(upper-left), (b) the narrow-width approximation with full spin
correlation (upper-right), and (c) it with diagonal spin correlation
(down-left) 
with the parameter set `gaugino region, light $\tilde{\nu}_e$' at
the CMS energy of 250~GeV. 
}
\end{figure}

This approximation is precise enough and much faster to reproduce the
cross section of the $exact$ calculation based on the full 54 diagrams
as one can see in Table.4. Furthermore, the distributions relevant to
the chargino spin and spin-spin correlations are also reproduced very
accurately as shown in Fig.~\ref{FIG:comparison} and
Fig.~\ref{FIG:phiphi}.
In the figures, `missing $p_T$' means the missing transverse momentum
carried by the neutralinos, `$p^{jet}_T$' the transverse momentum of
the quarks (parton level information of each quark), $\theta^*_-$ the
polar angle of the $s\bar{c}$ system with respect to the momentum of
$\tilde\chi^-$ in its rest frame, and $\phi^*_-$ its azimuthal angle
measured from the chargino pair-production plane.
Though $\phi^*_-$ is not an experimental observable, it is shown that
the event topology is correctly described by this approximation.
In the following sections we will present the results of the simulation 
study based on this approximation. 

Sometimes a similar approximation is used without the off-diagonal part 
in the spin summation as in~\cite{susy23}:
\begin{eqnarray}
\sigma&=&\frac{1}{C} \sum_{h_i} \int\!\!\!\int\!\!\!\int
\sum_{h+,h-}
{\left|{\cal A}_{e^+e^-\to\tilde\chi^+_1\tilde\chi^-_1}\right|}^2 
\cdot
{\left|{\cal A}_{\tilde\chi^+_1\to \tilde\chi^0_1 u \bar{d}}\right|}^2
\cdot
{\left|{\cal A}_{\tilde\chi^-_1\to \tilde\chi^0_1 s \bar{c}}\right|}^2
\d\Omega_{2\to2}
\nonumber\\
&~&\times
\frac{\d\Omega_{1\to3}}{2m_{\tilde\chi^\pm_1}\Gamma_{\tilde\chi^\pm_1}}
\frac{\d\Omega_{1\to3}}{2m_{\tilde\chi^\pm_1}\Gamma_{\tilde\chi^\pm_1}} 
\nonumber 
\end{eqnarray}
This approximation gives the same total cross sections as those with
full spin-correlation. However, it might give some deviation in
differential distributions, especially those of azimuthal angle
correlation between $\tilde\chi_1^+$ and $\tilde\chi_1^-$.
(See Figs.~\ref{FIG:comparison} and~\ref{FIG:phiphi}.)
The approximation with diagonal spin-correlation could be used for
limited studies, such as the detector acceptance estimation or the
measurements of the SUSY particle masses. It should, however, not be
used for the detailed studies of spins.

\section{Spin and spin-spin correlation measurements}
\subsection{Detector effects}
Even though there are two missing particles in the final state,
three angles can be obtained unambiguously from the experimental
observables as mentioned previously, $\cos\theta^*_\pm$, and
$\cos\Delta\phi^*$.
They are given by~\cite{tau}:
\begin{eqnarray}
\cos\theta^*_\pm&=&\frac{1}{\beta P_\pm}
\left(E_\pm-\frac{E^*_\pm}{\gamma}\right),
\nonumber \\
\sin\theta^*_+ \sin\theta^*_- \cos\Delta\phi &=&
\frac{P_+ P_-}{P^*_+ P^*_-}\cos\theta_{X^+X^-} + 
\frac{(E_+ - E^*_+/\gamma)(E_- - E^*_-/\gamma)}
{\beta^2P^*_{X^+} P^*_{X^-}}.
\nonumber 
\end{eqnarray}
The direct experimental observables are:
\begin{itemize}
\item $E_\pm$~: the energy of the $q\bar{q}'$ system in the laboratory 
frame,
\item $P_\pm=\sqrt{E_\pm^2-M_\pm^2}$~: the corresponding momentum,
where $M_\pm$ is the invariant mass of the $q\bar{q}'$ pair, 
is also an experimental observable,
\item $\theta_{X^+X^-}$~: the opening angle between two $q\bar{q}'$ 
systems in the laboratory frame.
\end{itemize}

On the other hand the following variables 
require the independent informations such as
the nominal beam-energy, the chargino and neutralino masses:
\begin{itemize}
\item $\beta,\gamma$~: the velocity and the Lorentz factor of 
the chargino calculated from the nominal beam energy and the chargino 
mass, 
\item $E^*_\pm$~: the energy of the $q\bar{q}'$ system in 
the rest frame of the mother chargino, which is calculated from 
the chargino and the neutralino masses and the measured $q\bar{q}'$ 
invariant mass,
\item $P^*_\pm$~: the corresponding momentum.
\end{itemize}

One cannot avoid the experimental errors in measuring the above
variables. There are two kinds of sources for these errors. The first
is the limited resolution of the detectors. To estimate the effect we
use {\tt PYTHIA}~\cite{pythia} to hadronize quarks and to make them to
decay into stable particles.
The obtained energies of the particles are smeared by an assumed
experimental resolution of the electro-magnetic (hadron) calorimeters:
these resolutions are supposed to be
$\sigma_E/\sqrt{E}=15\%/\sqrt{E}$ plus 1\% constant term
($\sigma_E/\sqrt{E}=40\%/\sqrt{E}$ plus 2\% constant term)
assuming a Gaussian distribution.
The second source stems from the incomplete knowledge of the colliding 
beam energies due to the initial state radiation (ISR) and the
beamstrahlung (BS). Though smearing of the beam energy does not affect
the energy measurement of the particles directly, this must give some
uncertainty to $\beta$ and $\gamma$ factors of the charginos and also
to the direction of the momentum of the charginos, which defines the
quantization axis of the chargino spin. Hence the chargino pair is not
necessarily produced in back-to-back configuration in the laboratory
frame. This means the quantization axis cannot be taken common for
both charginos in the laboratory frame. For the estimation of these
effects we use the simple structure function~\cite{sf} for ISR and the
utility {\tt Luminos}~\cite{luminos} for BS\@. The TESLA reference
parameters at the CMS energy of 500~GeV listed in Table.5 are used for
the BS simulation in the {\tt Luminos}.

\begin{figure}[htb]
\centerline{
\includegraphics[width=\textwidth]{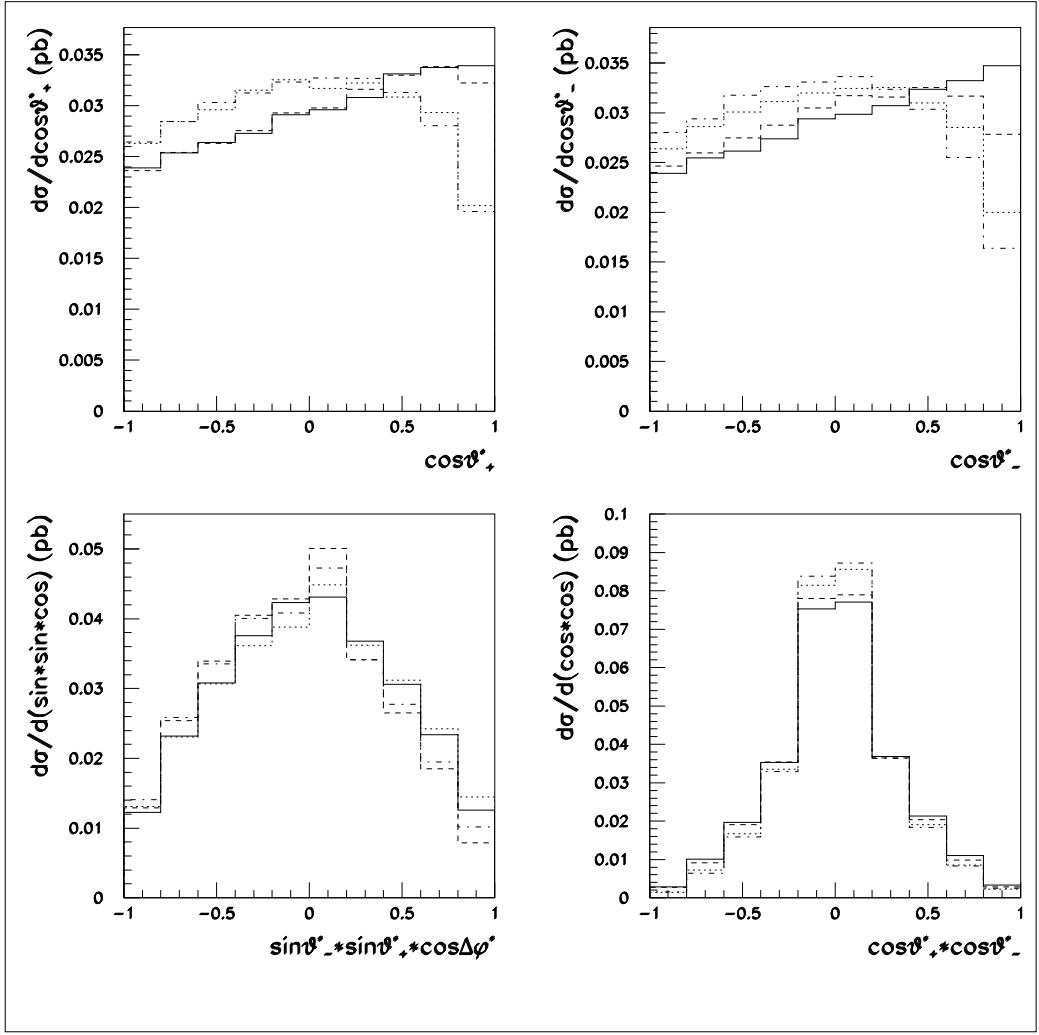}
}
\caption[FIG:deteff]{\label{FIG:deteff}
The effects of the detector resolution, ISR, and BS on the experimental
observable used in the spin measurements 
with the parameter set `gaugino region, light $\tilde{\nu}_e$' at
the CMS energy of 250~GeV. 
Solid lines show the original
distribution, dashed lines with smearing due to the limited 
resolution of
the calorimeters, dotted lines with ISR and BS, and dot-dashed lines 
with all experimental effects.
The parameter set of `gaugino region' with light $\tilde{\nu}_e$
is used.
}
\end{figure}

The calculated distributions of the experimental observable necessary
for the spin and spin-spin correlation measurements are shown in
Fig.~\ref{FIG:deteff}. The parameter set for the `gaugino region' with
a light $\tilde{\nu}_e$ mass is used for the benchmark.
The generated distributions are deformed due to the effects mentioned 
above as shown in Fig.~\ref{FIG:deteff}. 
One can see that the hadronization affects the heavy quark side
($\tilde\chi^0_1 s \bar{c}$) largely. The ISR and BS give more
deformation to the distributions than to the hadron energy
measurement. Even after including the experimental effects, one still
has the possibility to measure the spin and spin-spin correlations.

\subsection{Mixing angle dependence}
The cross section of the chargino pair-production depends on the
chargino mass, neutralino mass, the $\tilde{\nu}_e$ mass, and the
chargino mixing angles. The chargino and neutralino masses can be
determined from the threshold energy of the production and the energy
distribution of the quark pairs. In order to extract the rest of
parameters, the total cross section of the chargino pair-production
itself is helpful. The total cross section at the CMS energy of
250~GeV are summarized in Table.7 for the six different parameter
sets. However, the information is not enough to derive the other
unknown parameters such as the $\tilde{\nu}_e$ mass and the chargino
mixing angles.

As proposed in ref.~\cite{choi}, let us look at the spin and spin-spin 
correlations on the angular distributions at the CMS energy of 250~GeV.
There are two angular observables in the chargino spin measurement,
$\cos\theta^*_+$ and $\cos\theta^*_-$, and two observables for the
spin-spin correlation measurement, $\cos\theta^*_+ \cos\theta^*_-$
and $\sin\theta^*_+ \sin\theta^*_- \cos\Delta\phi$.

\begin{figure}[htb]
\centerline{
\includegraphics[width=\textwidth]{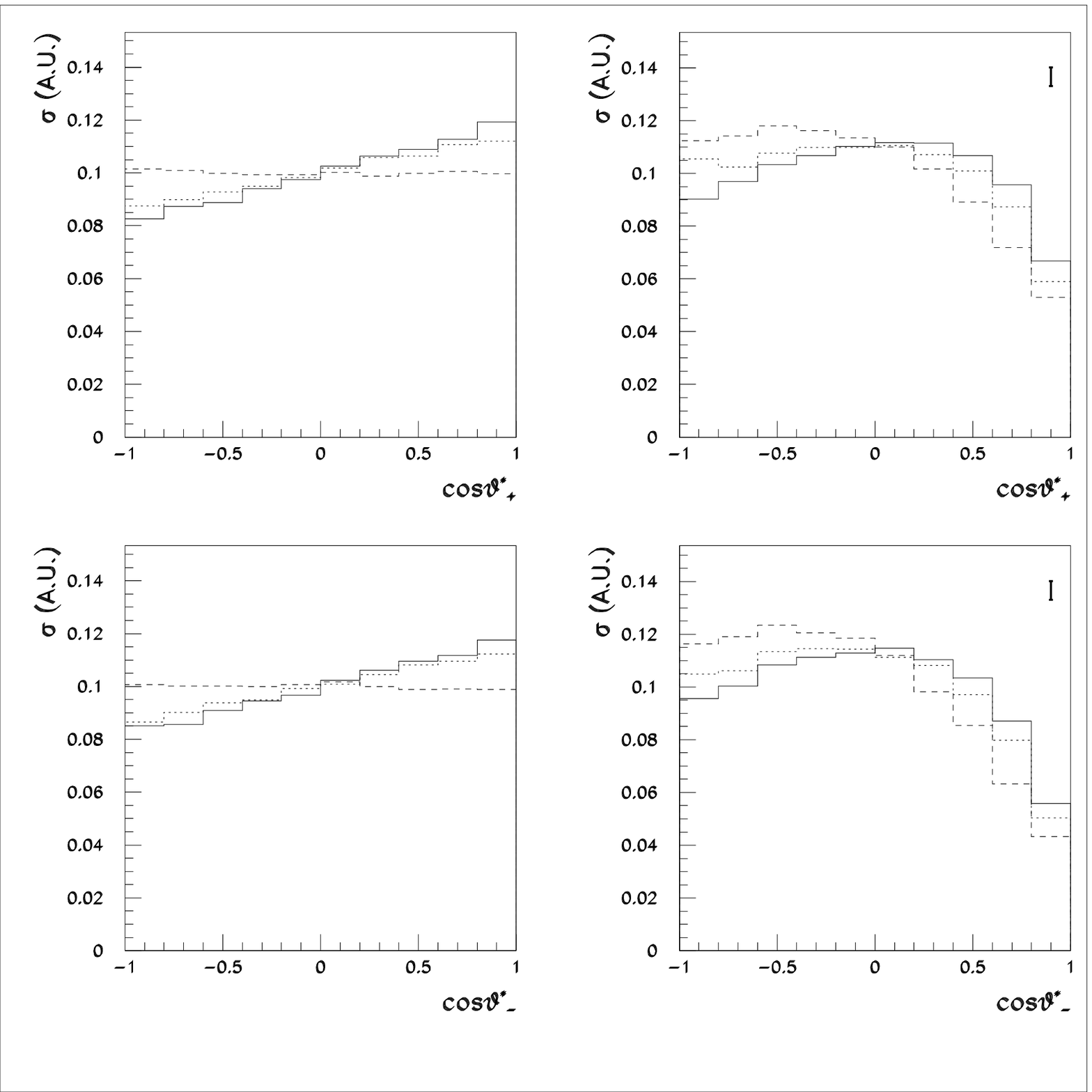}
}
\caption[FIG:light1]{\label{FIG:light1}
The distributions used for the chargino spin measurement
with parameters of the `light $\tilde{\nu}_e$ case'
at the CMS energy of 250~GeV.
The figures at left side show the original distribution 
and those at right side show those with the experimental effects.
Solid lines show those with parameters of `gaugino region',
dashed lines with `higgsino region', and dotted lines with
`mixed region'.
The error bars shown in the figures indicate the expected
statistical errors after accumulating a luminosity of 10~fb$^{-1}$.
All distributions are normalized to unity.
}
\end{figure}

\begin{figure}[htb]
\centerline{
\includegraphics[width=\textwidth]{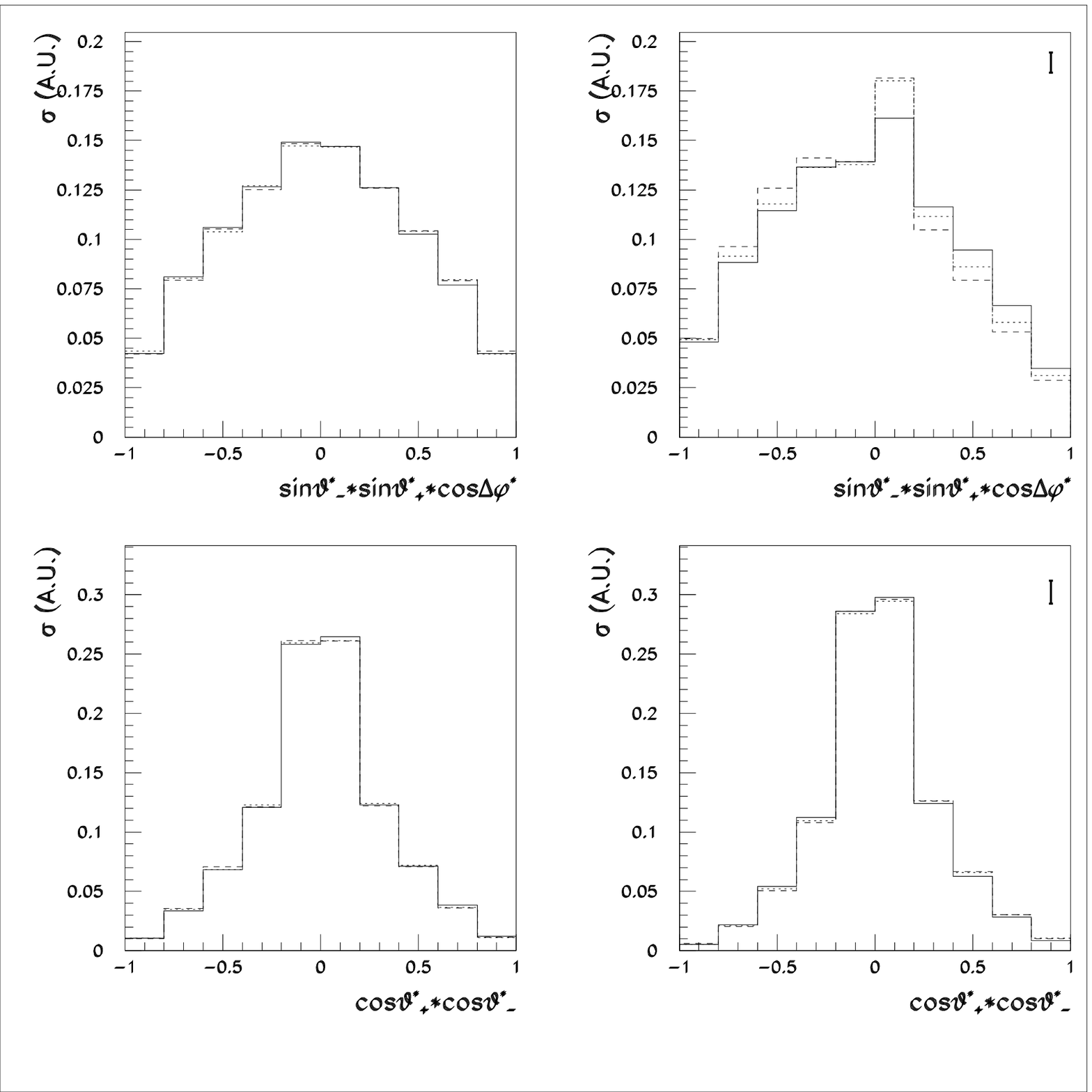}
}
\caption[FIG:light2]{\label{FIG:light2}
The distributions used for the chargino spin-spin correlation measurement
with parameters of the `light $\tilde{\nu}_e$ case'
at the CMS energy of 250~GeV.
The figures at left side show the original distribution 
and those at right side show those with the experimental effects.
Solid lines show those with parameters of `gaugino region',
dashed lines with `higgsino region', and dotted lines with
`mixed region'.
The error bars shown in the figures indicate the expected
statistical errors after accumulating a luminosity of 10~fb$^{-1}$.
All distributions are normalized to unity.
}
\end{figure}

\begin{figure}[htb]
\centerline{
\includegraphics[width=\textwidth]{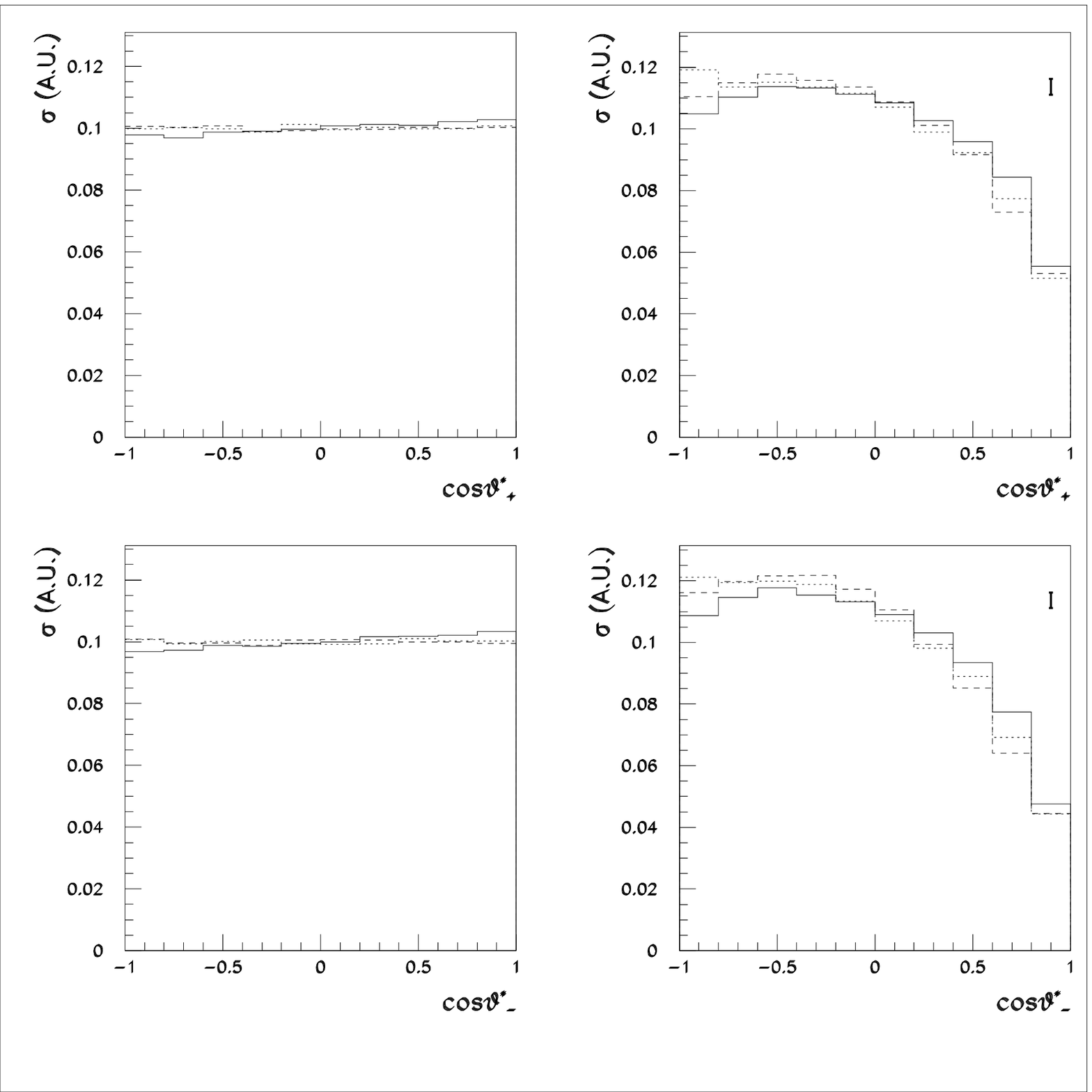}
}
\caption[FIG:heavy1]{\label{FIG:heavy1}
The distributions used for the chargino spin measurement
with parameters of the `heavy $\tilde{\nu}_e$ case'
at the CMS energy of 250~GeV.
The figures at left side show the original distribution 
and those at right side show those with the experimental effects.
Solid lines show those with parameters of `gaugino region',
dashed lines with `higgsino region', and dotted lines with
`mixed region'.
The error bars shown in the figures indicate the expected
statistical errors after accumulating a luminosity of 10~fb$^{-1}$.
All distributions are normalized to unity.
}
\end{figure}

First let us consider the light $\tilde{\nu}_e$ mass case. As shown 
in Fig.~\ref{FIG:light1} (the left column) the generated distribution
on $\cos\theta^*_\pm$ shows a clear difference among three
SUSY parameter sets, `gaugino region', `higgsino region', and
`mixed region'. 
(In order to see only the difference of the distributions but not
their absolute values, all distributions in 
Figs.~\ref{FIG:light1}-\ref{FIG:heavy1} are normalized to unity as
indicated by A.U. (arbitrary unit).)

It can be seen that the experimental effects modify the distribution to
a rather large extent. However, even with these experimental uncertainties, 
these three parameter regions can be distinguished as seen in
Fig.~\ref{FIG:light1} (the right column). The error bars in the figures 
show the expected (statistical) experimental error after accumulating
10~fb$^{-1}$ integrated luminosity\footnote{
All hadronic channels are summed up and 50\% of detection efficiency
is assumed.}. 
The limited knowledge of the beamstrahlung would give the biggest
uncertainty to the measurements. If the energy spectrum of the colliding 
beams are measured precisely enough, it will be able to measure 
the chargino spin and to distinguish three typical regions of 
the chargino mixing angles. On the other hand as shown in 
Fig.~\ref{FIG:light2} the distributions for the spin-spin correlation 
measurement overlap and cannot be separated even in the generator level 
without any smearing. 
The sensitivity to the spin-spin correlation is lower than to the spin
measurement itself. 
These sensitivity can be described in terms of analyzing-powers
appearing in the cross section formula: $\kappa_\pm(<1)$ is the spin
analyzing-power of $\tilde\chi^\pm$, and $\kappa_{c}=\kappa_+ \times
\kappa_-$ is the analyzing-power of the spin-spin correlation.
Then the analyzing power of the spin-spin correlation is smaller than
those of the spin measurements.

For a heavy $\tilde{\nu}_e$ mass case the situation is rather worse
than in the previous case. Even for the spin measurements the
sensitivity is not enough to distinguish three typical chargino
mixing-angle regions, as shown in Fig.~\ref{FIG:heavy1}.
In this case the contribution from the $\tilde{\nu}_e$ exchange
diagrams is greatly suppressed and this kills the sensitivity to
distinguish the gaugino and higgsino regions. 
The same studies at the CMS energy at 500~GeV have also been performed.
It is found that the situation does not change much from that at
250~GeV concerning the sensitivities for the measurement of the spin
and spin-spin correlations.

\section{Conclusions}
The total and the differential cross sections of the process, 
$e^+e^- \to \tilde\chi^+_1 \tilde\chi^-_1
\to (\tilde\chi^0_1 q \bar{q}')(\tilde\chi^0_1 q \bar{q}')$,
have been calculated numerically. By comparing the narrow width 
approximation with the $exact$ cross sections based on the numerical
helicity amplitudes with the full 54 diagrams, it was confirmed that
this approximation with full spin correlation,
could reproduce the details of the distributions very accurately.
The possibilities have been investigated to distinguish experimentally
three typical sets of the chargino-mixing angles by measuring the spin 
and spin-spin correlation. It was found that the light 
$\tilde{\nu}_e$ mass case could distinguish three cases (gaugino, 
higgsino and mixed region) in the chargino spin measurements once
the data of 10~fb$^{-1}$ is accumulated, even the distributions are 
largely distorted by the experimental effects. 
However, the measurements of the spin-spin correlation for the light
$\tilde{\nu}_e$ mass case and even the spin measurements for the heavy
$\tilde{\nu}_e$ mass case turned out to be difficult.

\vskip 1cm
The authors would like to thank Dr.~F.~Boudjema, Dr.~G.~B\'elanger,
Dr.~J.~Fujimoto, Prof.~M.~Kuroda and Prof.~Y.~Shimizu
for their fruitful discussions and suggestions. 
This work is supported in part by Ministry of Education, Science,
and Culture, Japan under Grant-in-Aid (No.08640388).
One of us (V.L.) is supported by a JSPS Fellowship (P97215).

\begin{table}[htbp]
\begin{center}
\begin{tabular}{|l||c|c|c|}
\hline
&\multicolumn{3}{c|}{light $\tilde{\nu}_e$ case} \\
\cline{2-4}{parameter}
&gaugino region&higgsino region&mixed region\\
\hline
\hline
tan$\beta$ & 2 & 2 & 2 \\
\hline
$\mu$ (GeV) & $-215.0$ & $-81.0$ & $-93.0$ \\
\hline
$M_2$ (GeV) & 81.4 & 215.0 & 92.0 \\
\hline
\hline
$\cos\phi_L$ & 0.914 & 0.056 & 0.437 \\
$\sin\phi_L$ & 0.405 & $-0.998$& 0.900 \\
\hline
$\cos\phi_R$ & 0.998 & 0.405 & 0.908 \\
$\sin\phi_R$ & 0.055 &$-0.914$ &$-0.419$ \\
\hline
\hline
$M_{\tilde{\chi_1}^\pm}$ (GeV) & 95.1 & 94.7 & 94.3 \\
\hline
$M_{\tilde{\chi_1}^0}$ (GeV) & 44.6 & 74.5 & 50.8 \\
\hline
$M_{\tilde{\nu}_e}$ (GeV) & 139.7& 140.0& 144.7\\
\hline
$M_{\tilde{\rm{U}}_{1(R)}}$ (GeV) & 284.9&689.2&315.2 \\ 
\hline
$M_{\tilde{\rm{U}}_{2(L)}}$ (GeV) & 279.2&668.6&308.3 \\
\hline
$M_{\tilde{\rm{D}}_{1(R)}}$ (GeV) & 291.5&691.9&321.2 \\ 
\hline
$M_{\tilde{\rm{D}}_{2(L)}}$ (GeV) & 280.4&666.9&309.1 \\ 
\hline
\end{tabular}
\end{center}
\caption{
Three parameter sets with a light $\tilde{\nu}_e$.
}
\end{table}
\begin{table}[htbp]
\begin{center}
\begin{tabular}{|l||c|c|c|}
\hline
&\multicolumn{3}{c|}{heavy $\tilde{\nu}_e$ case} \\
\cline{2-4}{parameter}
&gaugino region&higgsino region&mixed region\\
\hline
\hline
tan$\beta$  & 2 & 2 & 2 \\
\hline
$\mu$ (GeV) & $-215.0$ & $-$81.0 & $-$93.0 \\
\hline
$M_2$ (GeV) & 81.0 & 215.0 & 92.0 \\
\hline
\hline
$\cos\phi_L$ & 0.914 & 0.056 & 0.436 \\
$\sin\phi_L$ & 0.405 &$-$0.998 & 0.899 \\
\hline
$\cos\phi_R$ & 0.998 & 0.405 & 0.908 \\
$\sin\phi_R$ & 0.056 & $-$0.914&  $-$0.418\\
\hline
\hline
$M_{\tilde{\chi_1}^\pm}$ (GeV) & 94.7 & 94.7 & 94.3 \\
\hline
$M_{\tilde{\chi_1}^0}$ (GeV) & 44.3 & 74.5 & 74.5 \\
\hline
$M_{\tilde{\nu}_e}$ (GeV) & 403.2& 396.9& 396.9\\
\hline
$M_{\tilde{\rm{U}}_{1(R)}}$ (GeV) &284.5&680.1&392.6 \\ 
\hline
$M_{\tilde{\rm{U}}_{2(L)}}$ (GeV) &279.0&529.6&249.0 \\
\hline
$M_{\tilde{\rm{D}}_{1(R)}}$ (GeV) &291.2&765.9&487.9 \\ 
\hline
$M_{\tilde{\rm{D}}_{2(L)}}$ (GeV) &280.2&682.9&397.4 \\ 
\hline
\end{tabular}
\end{center}
\caption{
Three parameter sets with a heavy $\tilde{\nu}_e$.
}
\end{table}
\begin{table}[htbp]
\begin{center}
\begin{tabular}{ c |c|c|c|}
\cline{2-4}
&gaugino region&higgsino region&mixed region\\
\cline{2-4}
\hline
\multicolumn{1}{|c||}{light $\tilde{\nu}_e$ case }
& $3.00\times10^{-5}$ & $9.08\times10^{-6}$ & $2.35\times10^{-5}$ \\
\hline
\multicolumn{1}{|c||}{heavy $\tilde{\nu}_e$ case }
& $2.56\times10^{-5}$ & $8.96\times10^{-6}$ & $2.41\times10^{-5}$ \\
\hline
\end{tabular}
\end{center}
\caption{
The total width (in GeV unit) 
of chargino corresponding to the six sets of SUSY parameters.
}
\end{table}
\begin{table}[htbp]
\begin{center}
\begin{tabular}{|c|c|c|c|c|}
\hline
&\multicolumn{2}{c|}{6-body exact}&
		  \multicolumn{2}{c|}{Narrow-width Approx.} \\
\cline{2-5}{2-body$\times$Br.}&54 diag.&27 diag.& full-spin&diagonal \\
\hline
\hline
$58.68(1)$&$58.6(1)$&$58.6(1)$&
$58.66(6)$&$58.67(6)$ \\
\hline
\end{tabular}
\end{center}
\caption{
The total cross sections (in fb unit) of the process 
$e^+e^- \to \tilde\chi^0_1 u \bar{d} \tilde\chi^0_1 s \bar{c}$
with the parameter set `gaugino region, light $\tilde{\nu}_e$' at
the CMS energy of 250~GeV. 
The number in parenthesis is the statistical
error of the numerical integration on the last digit.
}
\end{table}
\begin{table}[htbp]
\begin{center}
\begin{tabular}{|r|c|}
\hline
\multicolumn{2}{|l|}{Beam parameters} \\
\hline
\hline
Number of particles & $2\times 10^{10}$/bunch \\
\hline
beam size $\sigma_x$ & 553nm \\
          $\sigma_y$ &   5nm \\
          $\sigma_z$ & 0.4mm \\
\hline
beam momentum spread & 1\%   \\
\hline
\end{tabular}
\end{center}
\caption{
The beam parameters used to calculate the beamstrahlung, taken from
the TESLA reference parameter set for the CMS energy of 500~GeV.
These parameters are used for the CMS energy of 250~GeV also.
}
\end{table}
\begin{table}[htbp]
\begin{center}
\begin{tabular}{ c |c|c|c|}
\cline{2-4}
&gaugino region&higgsino region&mixed region\\
\cline{2-4}
\hline
\multicolumn{1}{|c||}{light $\tilde{\nu}_e$ case }
& $58.74(8)$ & $141.3(2)$ & $65.90(9)$\\
\hline
\multicolumn{1}{|c||}{heavy $\tilde{\nu}_e$ case }
& $305.2(2)$ & $182.2(2)$ & $222.5(3)$\\
\hline
\end{tabular}
\end{center}
\caption{
The total cross sections (in fb unit) with the six parameter sets
including ISR and BS
with the parameter set `gaugino region, light $\tilde{\nu}_e$' at
the CMS energy of 250~GeV. 
The number in parenthesis is the statistical
error of the numerical integration on the last digit.
}
\end{table}


\begin{thebibliography}{99}

\bibitem{keisuke}T.~Tsukamoto, K.~Fujii, H.~Murayama, M.~Yamaguchi,
Y.~Okada, Phys.~Rev.~D {\bf 51} (1995) 3153.
\bibitem{tau}H.~K\"uhn, F.~Wagner, Nucl.~Phys. {\bf B236} (1984) 16. \\
G.~Goggi, Proc.~LEP summer study, CERN 79--01, 483.
\bibitem{gmp}
G.~Moortgat-Pick, H.~Fraas, A.~Bartl, W.~Majerotto, 
hep-ph/9804306. 
\bibitem{choi}S.Y.~Choi, A.~Djouadi, H.~Dreiner, J.~Kalinowsky, P.M.~Zerwas,
hep-ph/9806279. 
\bibitem{rge}V.~Lafage, in preparation.\\
For the program, please contact {\tt `lafage@minami.kek.jp'}.
\bibitem{GRACE}
T.~Ishikawa, T.~Kaneko, K.~Kato, S.~Kawabata, Y.~Shimizu, H.~Tanaka.
GRACE manual, KEK report 92-19, 1993.
\bibitem{GRACE-SUSY}Full Lagrangian of MSSM used in the GRACE system
is under preparation for publication. \\Please contact
{\tt `mkuroda@dave.hrz.uni-bielefeld.de'}.
\bibitem{BASES}
S.~Kawabata, Comp.~Phys.~Comm. {\bf 41} (1986) 127; {\it ibid.,\/} {\bf 88}
(1995) 309.
\bibitem{susy23}For example, {\bf susy23} is using this approximation:\\
J.~Fujimoto, K.~Hikasa, T.~Ishikawa, M.~Jimbo, T,~Kaneko, K.~Kato,
S.~Kawabata, T.~Kon, M.~Kuroda, Y.~Kurihara, T.~Munehisa, D.~Perret-Gallix,
Y.~Shimizu, H.~Tanaka, Comp.~Phys.~Commun. {\bf 109} (1998) 1.
\bibitem{pythia}T.~Sj\"ostrand, Comp.~Phys.~Comm. {\bf 82} (1994) 74.
\bibitem{sf}
E.~Kuraev, V.~Fadin, Yad.Phys., {\bf 41} (1985) 733 [Sov.J.Nucl.Phys.
41 (1985) 466]; \\
J.~Fujimoto, M.~Igarashi, N.~Nakazawa, Y.~Shimizu, K.~Tobimatsu,
Progr.Theor.Phys.Suppl,{\bf 100} (1990) 1.
\bibitem{luminos}N.~Toomi, J.~Fujimoto, S.~Kawabata, Y.~Kurihara,
T.~Watanabe, Phys.~Lett. {\bf B429} (1998) 162. \\
For the program, please contact {\tt `toomi@kekvax.kek.jp'}.

\end{thebibliography}
\end{document}